# Three-Dimensional Fermiology by Soft-X-Ray ARPES: Origin of Charge Density Waves in VSe$_2$

Vladimir N. Strocov[1][*], Ming Shi[1], Masaki Kobayashi[1], Claude Monney[1], Xiaoqiang Wang[1], Juraj Krempasky[1], Thorsten Schmitt[1], Luc Patthey[1], Helmuth Berger[2] and Peter Blaha[3]

[1]Swiss Light Source, Paul Scherrer Institute, CH-5232 Villigen-PSI, Switzerland
[2]Institut de Physique de la Matière Complexe, EPFL, CH-1015 Lausanne, Switzerland
[3]Institut für Materialchemie, Technische Universität Wien, A-1060 Wien, Austria

**Electronic structure of crystalline materials is their fundamental characteristic which is the basis of almost all their physical and chemical properties. Angle-resolved photoemission spectroscopy (ARPES) is the main experimental tool to study all electronic structure aspects with resolution in k-space. However, its application to three-dimensional (3D) materials suffers from a fundamental problem of ill-defined surface-perpendicular wavevector $k_z$. Here, we achieve sharp definition of $k_z$ to enable precise navigation in 3D k-space by pushing ARPES into the soft-X-ray photon energy range. Essential to break through the notorious problem of small photoexcitation cross-section was an advanced photon flux performance of our instrumentation. We explore the electronic structure of a transition metal dichalcogenide VSe$_2$ which develops charge density waves (CDWs) possessing exotic 3D character. We experimentally identify nesting of its 3D Fermi surface (FS) as the precursor for these CDWs. Our study demonstrates an immense potential of soft-X-ray ARPES (SX-ARPES) to resolve various aspects of 3D electronic structure.**

---

[*] Corresponding author. Adress: Swiss Light Source. Paul Scherrer Institute, CH-5232 Villigen-PSI, Switzerland. Email: vladimir.strocov@psi.ch. Telephone: +41 56 310 5311.

The momentum resolving ability of ARPES[1] is based on the fact that the photon promoted electron excitation in the crystal bulk, from an initial state in the valence band to final state in the unoccupied states continuum, conserves the 3D momentum **k**. Upon the following photoelectron escape into vacuum, the two-dimensional (2D) translational symmetry of the crystalline surface ensures conservation of the surface-parallel component $\mathbf{k}_{//}$ which can then be directly measured in the experiment. This makes ARPES the ideal tool for 2D materials. However, its application to 3D systems faces two fundamental difficulties on resolving the third surface-perpendicular $k_z$ component: (1) At the photoelectron escape stage, $k_z$ is distorted by the momentum absorbed by the surface barrier, and is no longer directly measurable. It can only be recovered if one knows the $E(k_z)$ dispersion of the final state back at the photoexcitation stage where the full 3D momentum was conserved. The free-electron (FE) approximation commonly used at this point has only a limited applicability, because for many materials the final states can feature complicated non-FE and excited-state self-energy effects[2,3]; (2) The final state confinement within the photoelectron escape depth $\lambda$ results, by the Heisenberg uncertainty principle, in intrinsic broadening of $k_z$ defined by $\Delta k_z = \lambda^{-1}$. The valence band $E(k_z)$ appearing in the measured spectrum is then an average over the $\Delta k_z$ interval, which intrinsically limits resolution of the ARPES experiment in $k_z$.[4]

Pushing ARPES from the conventional VUV to soft-X-ray photon energy range around 1 keV[5,6,7] addresses both problems: (1) The photoelectron energies become much larger than the crystal potential modulations, which makes the final states truly FE-like; (2) The increase of $\lambda$ with energy following the "universal curve"[8] results in sharpening of the $\Delta k_z$ intrinsic accuracy. Further advantages of SX-ARPES include larger bulk sensitivity and simplified matrix elements. However, SX-ARPES severely suffers from a loss of photoexcitation cross-section of valence states by a few orders of magnitude compared to the VUV energy range. We have solved this problem by using the high-resolution soft-X-ray beamline ADRESS[9] (ADvanced RESonant Spectroscopies) at Swiss Light Source (SLS) which delivers exceptionally high photon flux (see Methods).

Here, we apply SX-ARPES to explore 3D electronic structure of a typical layered transition metal dichalcogenide (TMDC) material $VSe_2$. A wide van der Waals gaps between the chalcogen-metal-chalcogen trilayers result in its quasi-2D properties[10,11]. The states derived from the out-of-plane orbitals like Se $4p_z$ retain nevertheless a 3D character with their $k_z$ dispersion range of a few eV. In VUV-ARPES, their response is distorted by non-FE final states and $\Delta k_z$ broadening comparable with the Brillouin zone (BZ) height $k_z^{BZ}$.[2,3]

Typical of the TMDCs are charge density waves (CDWs) appearing due to an interplay of the electron and phonon subsystems in the crystal[12,13,14] They may form, in particular, when large areas of the FS exhibit nesting with a **q**-vector vector leading to a soft phonon mode at this **q** freezing into the CDW. $VSe_2$ shows a CDW transition at $T_C \sim 110K$ accompanied by characteristic anomalies in transport and magnetic properties. Intriguingly, in contrast to most of the TMDCs[13,15,16], the CDWs in $VSe_2$ are 3D in the sense of their wavevector $\mathbf{q}^{CDW}$ having a large out-of-plane component[10,17]. The question is, are there precursors of such exotic 3D-CDWs in nesting properties of the FS?

We start with a general picture of the 3D electronic structure of $VSe_2$. In SX-ARPES experiment, one navigates in the 3D **k**-space by varying $\mathbf{k}_{//}$ through the emission angle $\vartheta$, and varying $k_z$ through photon energy $h\nu$ (with corrections for the photon momentum $p^{ph} = h\nu/c$)[18,19] From extremal behavior of the ARPES spectra with $h\nu$, we found $k_z$ to pass the Γ-point at 885 eV, see the BZ sketch in Fig. 1a. The corresponding ARPES intensity image $I(E,k_{//})$ along the M'ΓM line is shown in Fig. 1b. The $I(E,k_z)$ map along the surface-perpendicular ΓA direction in Fig. 1c was generated from images measured under $h\nu$ variation, with $h\nu$ rendered into $k_z$ assuming FE final states with an empirical inner potential of 7.5 eV. The experimental dispersions agree with the previous VUV-ARPES study[2] where control over $k_z$ was achieved by determination of the final states by Very-Low-Energy Electron Diffraction. Fig. 1d,e zoom in the Fermi level $E_F$ region along the M'ΓM and KΓK lines, respectively.

Statistics of our data is remarkable because at energies around 900 eV the V $3d$ and Se $4p$ cross-sections drop by a factor of ~1800 and 34, respectively, compared to a typical VUV-ARPES photon energy of 50 eV.[20] Profound contrast and dispersion of the spectral structures makes redundant any image enhancement like subtraction of angle-integrated background[21] or second-derivative representation[22]. This is remarkable because in the soft-X-ray range the photoelectron wavelength is comparable with amplitudes of the thermal motion of atoms, which acts to destroy the coherent spectral structures through their Debye-Waller amplitude reduction, broadening and piling up an incoherent **k**-integrated background[19,23]. Working at temperatures around 11 K, we minimized these destructive effects even for VSe$_2$ having rather low Debye temperature of 220 K.[24] Furthermore, excellent agreement of our Density Functional Theory (DFT) calculations (*blue lines* in Fig. 1) with the experiment demonstrates a weakly correlated nature of VSe$_2$. The calculations are seen however to underestimate the hybridization between the Se $4p_z^*$ and $4p_{xy}^*$ bands near the Γ-point.

Accurate control over $k_z$ in our experiment has allowed us to slice the FS in different planes. Fig. 2b shows the ΓALM slice acquired under $hv$ variation. Due to high kinetic energies, the iso-$hv$ trajectories (*dashed lines*) show only small $k_z$ variations with **k**$_{//}$. Fig. 2c,e display the ΓKM ($k_z$ = 0) and AHL ($k_z$ = ΓA) slices, respectively, and Fig. 2d a slice between them at $k_z \sim k_z^{BZ}/2$ acquired under variation of $\vartheta$. The definition of our experimental FS is well superior to previous VUV-ARPES results[22] which showed elliptical FS pockets from the AHL plane mixed with streaks near **k**$_{//}$ = 0 from the ΓKM plane.

Topology of the FS expected from our DFT calculations is shown in Fig. 2a, and its contours in the above BZ planes superimposed on the experimental data (*blue lines*). Again, the calculations show remarkable agreement with the experiment. For the ΓALM plane, the experimental FS exhibits somewhat larger area than the calculation suggesting a stronger out-of-plane conductivity. For the AHL slice, even fine details are reproduced such as a tiny rectangular distortion of the FS pockets around the L-points. For the ΓKM slice, the straight sections of the FS are well reproduced. The intense spot at the Γ-point comes from the intensity rise in the $4p_{xy}^*$ band just below $E_F$, see Fig. 1b, and streaks going from Γ in the K-direction follow the V 3$d$ band staying there flat near $E_F$, see Fig. 1e. We note no signs of the CDW superstructure in our data, indicating that the CDW induced perturbation of the crystal potential is weak enough to leave the spectral weight follow the fundamental periodicity of the VSe$_2$ lattice[25].

With confidence of our accurate navigation in 3D **k**-space, we turned to investigation of the electronic structure precursors of the 3D-CDWs in VSe$_2$. Theoretically, interaction of the electron and phonon systems in the crystal forms a CDW with wavevector **q** under the instability condition[14]

$$\frac{4\bar{\eta}_\mathbf{q}^2}{\hbar\omega_\mathbf{q}} \geq \frac{1}{\chi_\mathbf{q}} + 2\bar{U}_\mathbf{q} - \bar{V}_\mathbf{q}$$

where $\bar{\eta}_\mathbf{q}$ is the electron-phonon interaction corresponding to a phonon mode of energy $\omega_\mathbf{q}$, $\chi_\mathbf{q}$ is the (real part of) electronic susceptibility at the static $\omega \to 0$ limit, and $\bar{U}_q$ and $\bar{V}_q$ are matrix elements of their Coulomb and exchange interactions, respectively. This inequality to set up CDWs is driven either by increase of the left-hand "phononic" side due to strong electron-phonon interaction and availability of soft phonon modes[13,16], or by decrease on the right-hand "electronic" side due to singularity of $\chi_\mathbf{q}$ due to response of the conduction electrons coupled by the corresponding soft phonon mode. The latter scenario usually takes place when large parallel areas of the FS nest at certain **q** resulting in a peak of $\chi_\mathbf{q}$.[13,15] In contrast to the one-dimensional (1D) case automatically leading to the paradigm Peierls transition, realization of this scenario in 3D is most restrictive because the nesting areas should match their curvatures in all three dimensions. We will illustrate how this is realized in VSe$_2$.

The 3D-CDWs are characterized by $\mathbf{q}^{CDW} = q_{//}^{CDW}\mathbf{a} + q_z^{CDW}\mathbf{c}$, where **a** and **c** are the in-plane and out-of-plane reciprocal lattice unit vectors, respectively. The actual values of $q_{//}^{CDW}$ and $q_z^{CDW}$ in VSe$_2$ determined by X-ray diffraction are about 0.25$a$ (~0.54 Å$^{-1}$) and 0.33$c$ (~0.34 Å$^{-1}$) respectively, with some scatter in the literature regarding their exact values, commensurability and transition temperatures[17,26]. VUV-ARPES studies of VSe$_2$ have related the $q_{//}^{CDW}$ component to nesting of the FS along the **a**-axis[22,27] and even identified a weak spectral weight reduction in the corresponding **k**$_{//}$ regions[22]. According to theoretical analysis[10], the $q_z^{CDW}$ one would originate from nesting along the **c**-axis owing to warping of the FS in the $k_z$ direction. However, VUV-ARPES failed to yield any unambiguous evidence of this[28].

To resolve the origin of $q_z^{CDW}$, we have measured the out-of-plane FS cut in the MLL'M' plane, Fig. 3a. This plane goes through the region of (flattened by the rectangular distortion) FS areas which nest with $q_{//}^{CDW}$, see Fig. 2c. One of the $I(E,k_{//})$ images, measured near the MM' line at $k_z = 0$, is shown in Fig. 3b. The experimental FS cut obtained under $hv$ variations is shown in Fig. 3c. The cut shows 3D warping with concave and convex regions, which clearly nest through 3D **q**-vector (*arrows*) closely matching that of the 3D-CDWs.

We have confirmed the observed nesting by calculations of autocorrelation[15] $R(\mathbf{q}) = \int_\Omega I_F(\mathbf{k}) I_F(\mathbf{k}+\mathbf{q}) d^2\mathbf{k}$ of the (regularized) experimental FS map $I_F(\mathbf{k})$ throughout the BZ cut $\Omega$. The $R(\mathbf{q})$ map in Fig. 3d shows a gross arc-like peak identifying the nesting vector. As $R(\mathbf{q})$ is less sensitive to sliding of the bands in the $k_z$ direction, the peak is elongated in $q_z$. In principle, the $R(\mathbf{q})$ structures are only connected with $\text{Im}\chi_\mathbf{q}$ and might be distorted when carrying them over to $\text{Re}\chi_\mathbf{q}$ to define **q** of the CDW instability[13]. Despite these potential complications, the experimental $R(\mathbf{q})$ closely matches the actual 3D-CDWs: The $R(\mathbf{q})$ peak is centered at $q_{//} \sim 0.24a$ and in the $q_z$ direction extends from $\sim 0.3$ to $0.5c$. An energy gain under commensuration of the CDW[12] slightly shifts the system along the $R(\mathbf{q})$ arc to the nearest commensurate **q** which has the actual $q_{//}^{CDW} \sim 0.25a$ and $q_z^{CDW} \sim 0.33c$.

The formation of 3D-CDWs in VSe$_2$ follows therefore the FS nesting scenario, with their 3D character resulting from the 3D warping of the FS. We note that formation of 3D-CDWs is a rare phenomenon observed, to the best of our knowledge, only for two other TMDCs, TaS$_2$ and TiSe$_2$.[10,29] Our picture of the electronic structure precursors of the 3D-CDWs calls for a complementary study on the phononic side to identify the corresponding soft phonon mode. While Raman scattering experiments on VSe$_2$ deliver information[26] only on phonons in the $\Gamma$-point, inelastic X-ray scattering[16] is capable of probing any region of **q**-space.

Our ARPES results remain so far unique to detail the FS of VSe$_2$, because alternative experimental methods based on quantum oscillations (like de Haas – van Alphen effect) require large crystals of high purity hardly available for TMDCs. We note that the advantages of SX-ARPES to deliver sharp definition of $k_z$, combined with advanced photon flux performance of our instrumentation, have resulted in an exceptional clarity of our experimental $E(\mathbf{k})$ and FS maps. This has been crucial to unambiguously identify the precursors of the exotic 3D-CDWs in VSe$_2$.

## Methods

The experiments have been performed at the ADRESS beamline[9] of SLS. This beamline delivers soft-X-ray radiation with variable linear and circular polarizations in the energy range from 300 to 1600 eV to the RIXS and ARPES endstations. Its key feature is high photon flux up to $10^{13}$ photons/s/0.01%BW at 1 keV which has allowed us to a break through the notorious problem of small valence band crossection in the soft-X-ray region. The ARPES endstation uses a vertical scattering plane geometry (see Supplementary Information) with a grazing light incidence angle around 20º to increase the photoyield. The CARVING manipulator provides with three angular degrees of freedom. The sample environment allows cooling down to 10.7 K to quench the electron-phonon interaction effects destructive for the **k**-resolution. The photoelectron analyzer is PHOIBOS-150 normally operated with an acceptance of ±6º sufficient to cover more than one BZ. The measurements are normally performed at a combined beamline and analyzer $\Delta E$ of ~120 meV around $hv$ = 900 eV, delivering ARPES images of publication quality within a few minutes. For high-resolution measurements we sharpen $\Delta E$ to ~ 70 meV, with the acquisition time increasing to a few tens of minutes. In the present experiments, we used *p*-polarization of incident light which delivered stronger intensity compared to *s*-polarization, in particular for the V 3*d* bands. The analyser slit was oriented perpendicular to the scattering plane.

The band structure computations were performed within the standard DFT formalism. The electron exchange-correlation was described within the Generalized Gradient Approximation (GGA). The calculations employed a full-potential (Linearized) Augmented Plane Waves + Local Orbitals method implemented in the WIEN2k package[30].

## Acknowledgements

We thank J.H. Dil and V. Zabolotnyy for valuable advices, E.E. Krasovskii and J. Minár for critical reading of the manuscript, and C. Quitmann, J.F. van der Veen and J. Mesot for their continuous support of the SX-ARPES project at SLS. P.B. was supported by the Austrian Science Fund (SFB F41, "ViCoM").

**References**

1. Hüfner, S. Photoelectron Spectroscopy: Principles and Applications (Springer, 1996).
2. Strocov, V.N. *et al*. New method for absolute band structure determination by combining photoemission with very-low-energy electron diffraction: Application to layered $VSe_2$. *Phys. Rev. Lett.* **79**, 467-470 (1997).
3. Strocov, V.N. *et al*. Three-dimensional band structure of layered $TiTe_2$: Photoemission final-state effects. *Phys. Rev. B* **74**, 195125 (2006).
4. Strocov, V.N. Intrinsic accuracy in 3-dimensional photoemission band mapping. *J. Electron Spectrosc. Relat. Phenom.* **130**, 65-78 (2003).
5. Suga, S. *et al.* High-energy angle-resolved photoemission spectroscopy probing bulk correlated electronic states in quasi-one-dimensional $V_6O_{13}$ and $SrCuO_2$. *Phys. Rev. B* **70**, 155106 (2004).
6. Mulazzi, M. *et al*. Absence of nesting in the charge-density-wave system $1T-VS_2$ as seen by photoelectron spectroscopy. *Phys. Rev. B* **82**, 075130 (2010).
7. Sekiyama, A. *et al*. Technique for bulk Fermiology by photoemission applied to layered ruthenates. *Phys. Rev. B* **70**, 060506(R) (2004).
8. Powell, C.J. *et al*. Surface sensitivity of Auger-electron spectroscopy and X-ray photoelectron spectroscopy. *J. Electron Spectrosc. Relat. Phenom.* **98/99** (1999) 1-15
9. Strocov, V.N. *et al*. High-resolution soft X-ray beamline ADRESS at the Swiss Light Source for resonant inelastic X-ray scattering and angle-resolved photoelectron spectroscopies. *J. Synchrotron Rad.* **17**, 631-643 (2010).
10. Woolley, A.M. & Wexler, G. Band structures and Fermi surfaces for $1T-TaS_2$, $1T-TaSe_2$ and $1T-VSe_2$. *J. Phys. C: Solid State Phys.* **10**, 2601-2616 (1977).
11. Starnberg, H.I. *et al*. 3D-to-2D transition by Cs intercalation of $VSe_2$. *Phys. Rev. Lett.* **70**, 3111 (1993).
12. Gruener, G. *Density Waves in Solids* (Addison-Wesley, Reading, MA, 1994).
13. Johannes, M.D. & Mazin, I.I. Fermi surface nesting and the origin of charge density waves in metals. *Phys. Rev. B* **77**, 165135 (2008).
14. Chan, S.K. & Heine, V. Spin density wave and soft phonon mode from nesting Fermi surfaces. *J. Phys. F* **3**, 795-809 (1973).
15. Inosov, D.S. *et al*. Fermi surface nesting in several transition metal dichalcogenides. *New J. of Phys.* **10**, 125027 (2008),
16. Weber, F. *et al*. Extended phonon collapse and the origin of the charge-density wave in $2H-NbSe_2$. *Phys. Rev. Lett.* **107**, 107403 (2011),
17. Eaglesham, D.J., Withers, R.L. & Bird, D.M. Charge-density-wave transitions in $1T-VSe_2$. *J. Phys. C: Solid State Phys.* **19**, 359-367 (1986).
18. Venturini, F. *et al*. Soft x-ray angle-resolved photoemission spectroscopy on Ag(001): Band mapping, photon momentum effects, and circular dichroism. *Phys. Rev. B* **77**, 045126 (2008).
19. Gray, A.X. *et al*. Probing bulk electronic structure with hard X-ray angle-resolved photoemission. *Nature Materials* **10**, 759-764 (2011).
20. Yeh J.J. & Lindau, I. Atomic subshell photoionization cross sections and asymmetry parameters: $1<Z<103$. *Atomic Data and Nuclear Data Tables*, **32** (1985) 1-155 (Web-version available at http://ulisse.elettra.trieste.it/services/elements/WebElements.html)
21. Önsten, A. *et al*. Probing the valence band structure of $Cu_2O$ using high-energy angle-resolved photoelectron spectroscopy. *Phys. Rev. B* **76**, 115127 (2007).
22. Terashima, K. *et al*. Charge-density wave transition of $1T-VSe_2$ studied by angle-resolved photoemission spectroscopy. *Phys. Rev. B* **68**, 155108 (2003).
23. Plucinski, L. *et al*. Band mapping in higher-energy x-ray photoemission: Phonon effects and comparison to one-step theory. *Phys. Rev. B* **78** (2008), 035108
24. Kamarchuk, G.V. *et al*. Direct determination of Debye temperature and electron-phonon interaction in $1T-VSe_2$. *Phys. Rev. B* **63**, 073107 (2001).
25. Voit, J. *et al*. Electronic structure of solids with competing periodic potentials. *Science* **290**, 501-503 (2000).
26. Sugai, S. Lattice vibrations in the charge-density-wave states of layered transition metal dichalcogenides. *phys. stat. sol. (b)* **129**, 13-39 (1985).
27. Hughes, H.P., Webb, C. & Williams, P.M. Angle-resolved photoemission from $VSe_2$. *J. Phys. C: Solid St. Phys.* **13**, 1125-1138 (1980).
28. Sato, T. Three-dimensional Fermi-surface nesting in $1T-VSe_2$ studied by angle-resolved photoemission spectroscopy. *J. Phys. Soc. Jpn.* **73**, 3331-3334 (2004).
29. Monney, C. *et al*. Probing the exciton condensate phase in $1T-TiSe_2$ with photoemission. *New J. of Phys.* **12**, 125019 (2010).
30. P. Blaha, K. Schwarz, G. K. H. Madsen, D. Kvasnicka, and J. Luitz, *WIEN2k, An Augmented Plane Wave Plus Local Orbitals Program for Calculating Crystal Properties* (Vienna University of Technology, Austria, 2001).

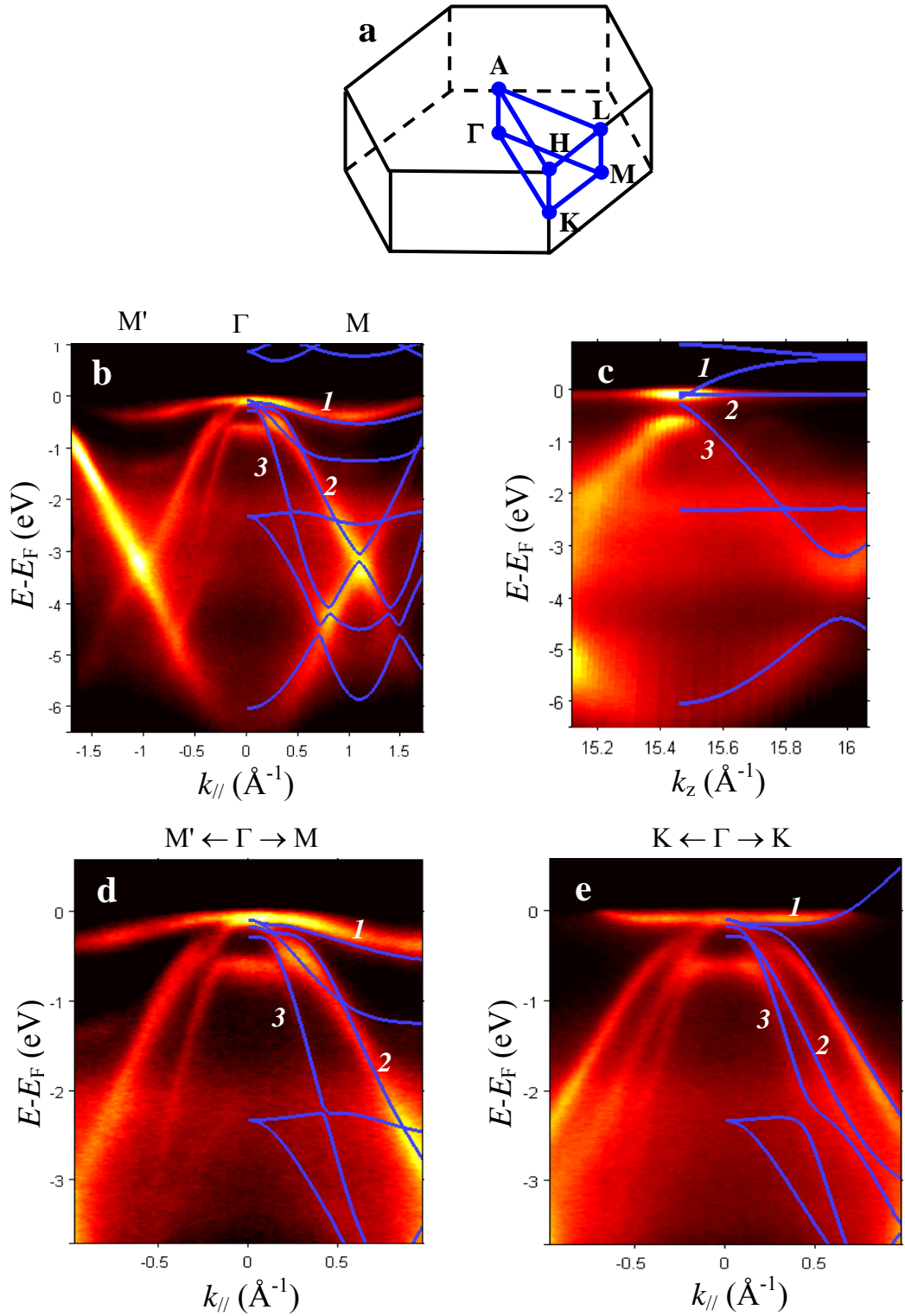

**Figure 1. SX-ARPES experimental band structure of VSe$_2$ along selected BZ lines. a**, BZ sketch; **b-e**, experimental ARPES intensity in colorscale: **b**, $I(E,k_{//})$ image along the M'ΓM line ($hv$ = 885 eV to deliver $k_z$ = 0). The energy resolution $\Delta E$ was ~120 meV and acquisition time only 7 min; **c**, $I(E,k_z)$ map along the ΓA line ($hv$ variation from 845 to 960 eV). Note different $k_z$ and $k_{//}$ scales in the figures. To compensate the cross-section variations, the individual $hv$ = $const$ slices are normalized in integral intensity; **d-e**, high-resolution $I(E,k_{//})$ images along the M'ΓM and KΓK lines ($hv$ = 885 eV), with $\Delta E$ sharpened here ~70 meV acquisition time increased to 40 min. One clearly observes the V 3$d$ bands (marked *1*) to form the FS, and the Se 4$p_{xy}^*$ (*2*) and 4$p_z^*$ (*3*) bands deeper in the valence band. The DFT band calculations (*blue lines*) remarkably agree with the experiment except underestimation of the Se 4$p_z^*$– $p_{xy}^*$ hybridization near the Γ-point.

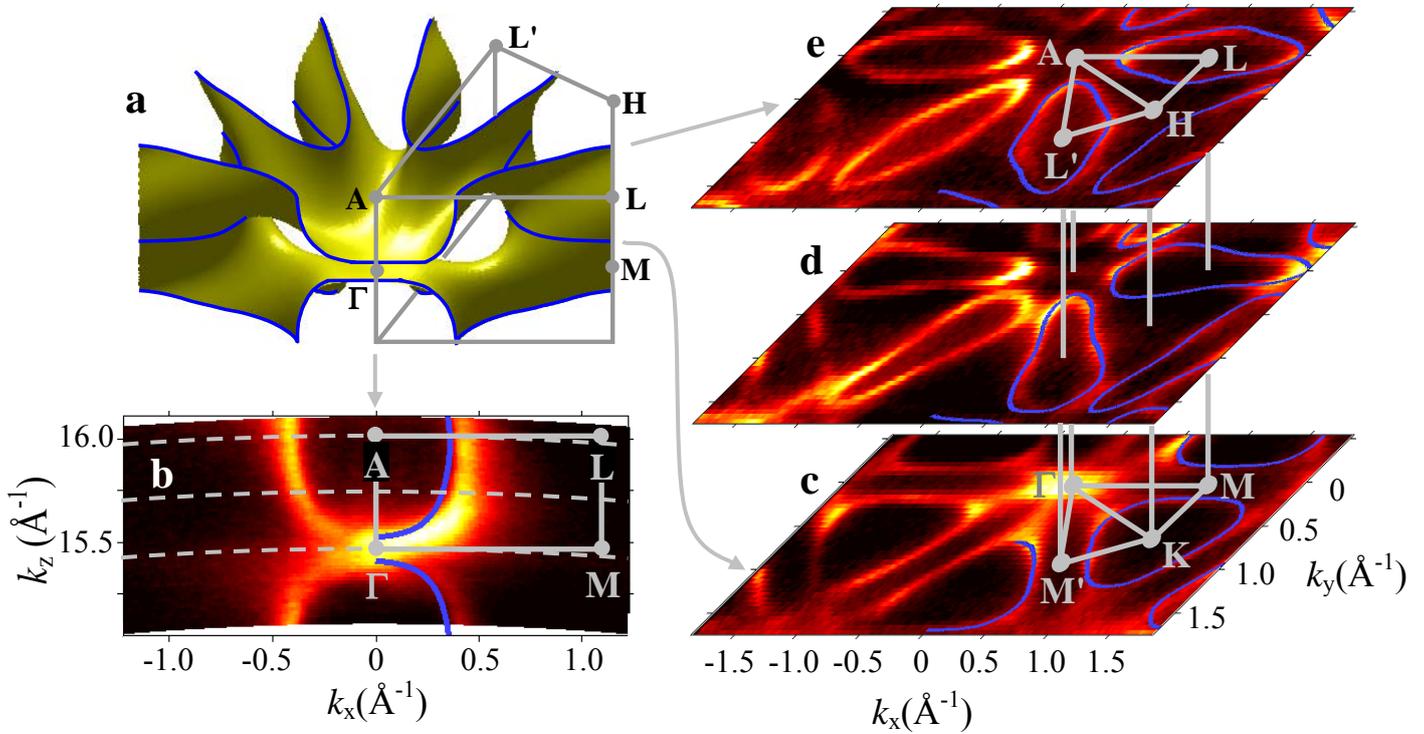

**Figure 2. Experimental FS slices in selected BZ planes. a**, DFT calculated FS of VSe$_2$ with its contours in the symmetry planes (*blue lines*); **b**, experimental FS slice in the ΓALM plane of the BZ ($h\nu$ = 845 to 960 eV). *Dashed lines* are iso-$h\nu$ trajectories; **c-e**, experimental FS slices in the ΓKM central plane ($h\nu$ = 885 eV), a plane at half the BZ heights (915 eV) and ALH face plane (945 eV), respectively. No image enhancement or symmetrization has been applied to these maps. The ΓKM and ALH maps follow the 6-fold symmetry characteristic of these symmetry planes, and the map between them displays asymmetric dog-bone electron pockets following the 3-fold symmetry of the BZ interior. By virtue of well-defined $k_z$ the experimental FS demonstrates a textbook clarity. The DFT calculations (*blue lines*) agree with the experiment, except for the central spot and streaks in the ΓKM plane coming from the V 3$d$ and Se 4$p_{xy}^*$ bands in vicinity of $E_F$.

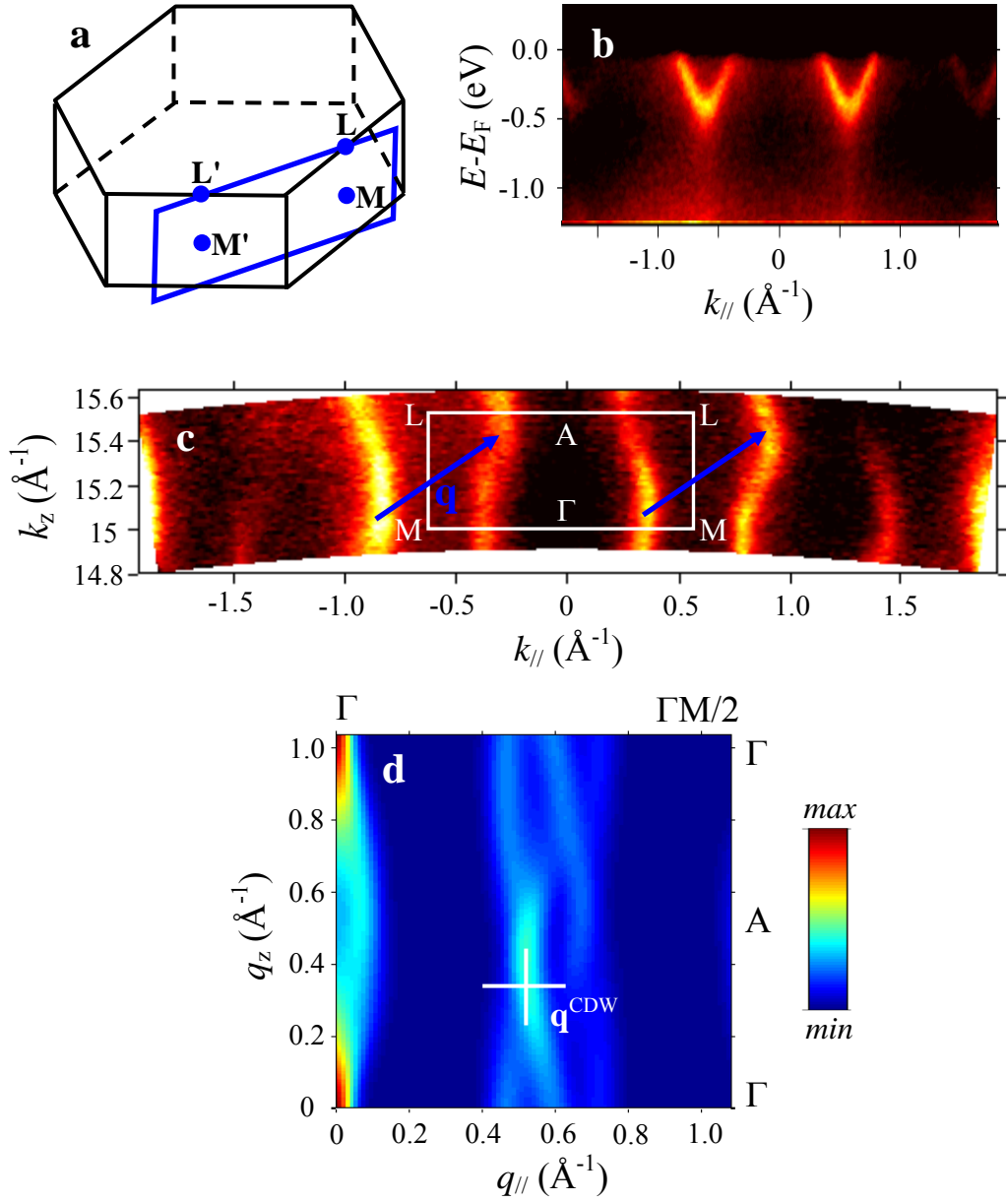

**Figure 3. 3D warping of the FS in the nesting region as the 3D-CDWs precursor. a**, MLL'M' plane of the BZ going through the nesting region; **b**, Experimental $I(E,k_{//})$ image along the MM' line ($hv$ = 890 eV); **c**, Experimental out-of-plane FS cut in the MLL'M' plane ($hv$ = 880 to 960 eV) normalized similar to Fig. 1c. The $k_z$ dispersions are asymmetric due to the 3-fold symmetry of the BZ interior. 3D warping of the FS contours results in nesting with the indicated **q** close to $\mathbf{q}^{CDW}$ of the 3D-CDWs; **d**, Corresponding $R(\mathbf{q})$ autocorrelation map showing an arc-like maximum near $\mathbf{q}^{CDW}$.

## Supplementary Information

## Experimental geometry

Geometry of our SX-ARPES experiment is sketched in the figure. The grazing light incidence angle, corresponding to the normal emission sample position, is $20^o$. Compared to the standard angle of $45^o$, this increases the photoyield by a factor of about 2. The sample is rotated towards the grazing incidence around the horizontal axis to balance the larger horizontal and smaller vertical light footprints. The spot size on the sample is confined within $30\times75$ μm$^2$ for all sample rotations. The manipulator primary rotation axis is horizontal and perpendicular to the vertical scattering plane formed by the incident light and the surface normal. The analyzer lens axis lies in the vertical scattering plane. The analyser can be turned around the lens to orient its slit either in the scattering plane, which allows exploring symmetries of the valence states by switching the incident light polarization, or perpendicular to it. In the first case we change 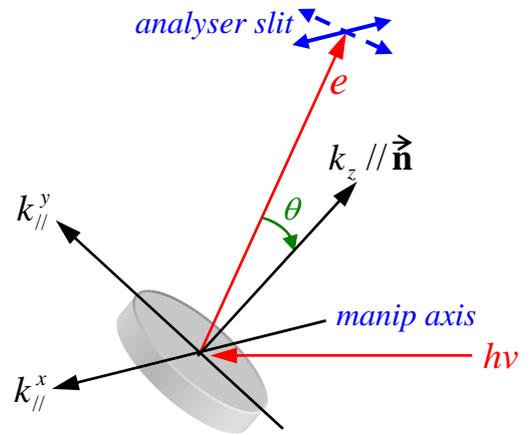 of the emission angle by tilt rotation of the sample, and in the second by primary rotation of the manipulator. The present experiment used the latter geometry. Our data acquisition software allows synchronization of the sample rotation angles with $h\nu$ to track the corresponding changes in photon momentum and photoelectron $\mathbf{k}_{//}$.